\documentclass[floatfix,amsmath,amssymb,prl,aps,showpacs,twocolumn,10pt]{revtex4-1}
\usepackage{amsmath,amsfonts,amssymb,amsthm,graphics,graphicx,epsfig,bbm}
\usepackage[colorlinks=true,citecolor=blue,linkcolor=red,urlcolor=blue]{hyperref}
\usepackage[usenames]{color}
\usepackage{graphicx}
\usepackage{subfigure}
\usepackage{amsmath}
\usepackage{epsfig}
\usepackage{dcolumn}
\usepackage{bm}
\usepackage{caption}
\usepackage{color}
\usepackage{epstopdf}
\usepackage{amssymb}
\usepackage{amstext}
\usepackage{latexsym}
\usepackage{hyperref}
\usepackage{amsfonts}
\usepackage{psfrag}
\usepackage{xcolor}
\usepackage[normalem]{ulem}
\usepackage{dsfont}
\usepackage{txfonts}
\usepackage{overpic}
\usepackage{ifthen}
\usepackage{amsthm}
\usepackage{float}
\usepackage{dblfloatfix}

\usepackage{amsmath, amssymb, amsthm}

\renewcommand{\Re}{\mathrm{Re}}

\newcommand{\ket}[1]{\vert #1 \rangle}
\newcommand{\bra}[1]{\langle #1 \vert}

\newcommand{\an}[2]{\ifthenelse{\equal{#1}{}}{\ensuremath{\hat{#1}_{#2}}}{\ensuremath{\hat{#1}^{\protect\phantom{\dagger}}_{#2}}}}

\begin{document}

\title{Out-of-time-order correlators in electronic structure using Quantum Computers}

\author{K. J. Joven$^{1}$}
\altaffiliation[]{These authors contributed equally to this work.}
\email{kevin.joven@ntt-research.com}
%%%
\author{V. M. Bastidas$^{1,2}$}
\altaffiliation[]{These authors contributed equally to this work.}
\email{victor.bastidas@ntt-research.com}
%%%
\affiliation{%
$^1$Physics and Informatics Laboratory, NTT Research, Inc., 940 Stewart Dr., Sunnyvale, California, 94085, USA
} 
\affiliation{%$
$^2$Department of Chemistry, Massachusetts Institute of Technology, Cambridge, Massachusetts 02139, USA
} 
\date{\today}

\begin{abstract}
    Operator spreading has profound implications in diverse fields ranging from statistical mechanics and blackhole physics to quantum information. The usual way to quantify it is through out-of-time-order correlators (OTOCs), which are the quantum analog to Lyapunov exponents in classical chaotic dynamics. In this work we explore the phenomenon of operator spreading in quantum simulation of electronic structure in quantum computers. To substantiate our results, we focus on a hydrogen chain $H_4$ and demonstrate that operator spreading is enhanced when the chain is far from its equilibrium geometry.  
    We also investigate the dynamics of bipartite entanglement and its dependence on the partition's size. Our findings reveal distinctive signatures closely resembling area- and volume-laws in equilibrium and far-from-equilibrium geometries, respectively.  Our results provide insight of operator spreading of coherent errors in quantum simulation of electronic structure and can be experimentally implemented in various platforms available today.
\end{abstract}

\maketitle

Chemistry offers a plethora of problems where complex dynamics can naturally appear~\cite{Kohn1999,whitfield2013,McArdle2020}. External fields or perturbations applied to molecules can induce electronic transitions between different potential surfaces and generate complex molecular dynamics~\cite{Deumens1994,Lisinetskaya2011}. Even under the Born-Oppenheimer approximation~\cite{Born1927}, due to the huge number of electronic degrees of freedom, the problem of electronic structure becomes intractable using classical computers~\cite{OGorman2022}. One alternative to overcome this problem is to use quantum devices to simulate it~\cite{quantumchemistrytoday,perturbations,Kassal2011}, which is one of the most promising near-term applications of quantum computers~\cite{ Bharti2022,Kim2023}. In fact, with currently existing technologies,  
we have access to quantum devices that already can solve some small-size problems in quantum chemistry including electronic structure and molecular vibrations~\cite{huh2015,quantumchemistrytoday,kandala2017,kandala2018,google2020}. One of the challenges of quantum simulation is to understand propagation of coherent or incoherent errors~\cite{heyl2019,kuper2022,GonzalesGarcia2022} and how to correct them~\cite{devitt2013,Choi2020}. 

 In a typical quantum simulation, coherent errors appearing during the computation can be treated as local unitaries acting on our system~\cite{Kueng2016,Suzuki2017,Huang2019}. The aim of this work is to investigate operator spreading of these unitaries during a quantum simulation of a molecule in equilibrium and far-from-equilibrium geometries. To substantiate our ideas, we consider a one-dimensional hydrogen chain $\text{H}_4$. We encode the electronic structure Hamiltonian in terms of qubits thus creating a quantum circuit with a parametric dependence on the geometry of the chain. We explore how local operators propagate in the system using out-time-order correlators (OTOCs)~\cite{LS_echo,operator_spreading,garttner2017}, a common tool used to investigate quantum signatures of chaos~\cite{GarciaMata2018} and information scrambling\cite{Shen2020}.

To achieve this, we leverage tools like the Jordan-Wigner encoding~\cite{jordan1928,tranter2018} of the electronic Hamiltonian for $\text{H}_4$ in the minimal basis using $8$ qubits. Motivated by Ref.~\cite{Mi2021} we propose a protocol to measure OTOCs in quantum simulation of electronic structure.  We first propagate a separable state forwards in time during time $T$, then apply a local orbital rotation to a given qubit and afterwards we propagate backwards in time during the same time $T$. In this way, the measurement of a local Pauli operator in the resulting state give us the OTOC. Further, we calculate the dynamics of bipartite entanglement and investigate how it depends on the partition size for different molecular geometries.

Recent works exploring OTOcs in chemistry have investigated information scrambling in a model of chemical reaction based on a double-well reaction coordinate~\cite{zhang2024}, in vibrational energy hopping~\cite{Zhang2022,Li2023}, and in ring-polymer molecular dynamics for a classically chaotic double-well model~\cite{Sadhasivam2023}.
In our work we decided to take a different route to previous works and instead of exploring continuous degrees of freedom, we tackle the problem of operator spreading and entanglement in quantum simulations of electronic structure, which involves discrete degrees of freedom represented using qubits. We demonstrate that when simulating electronic structure in equilibrium geometries, local operators remain bounded in the system due to the small entanglement created during quantum dynamics. In stark contrast to this, far from the equilibrium geometry there is strong operator spreading as the bipartite entanglement grows proportionally to the partition's size, thus resembling a volume law of entanglement.

The field of quantum simulation of dynamics for quantum chemistry in quantum computers is rather unexplored~\cite{miessen2023}. Our work contributes in understanding how molecular geometry relates to operator spreading and entanglement, which is relevant for near-term implementations of electronic structure using quantum computers and for method developments in quantum chemistry.

\textit{Electronic structure in quantum computers:}
The understanding of electronic structure stands as the most challenging problem in chemistry as it requires to numerically resolve the interactions between atoms and nuclei within specific configurations~\cite{Johnson1975,Kohn1999,shao2015advances,Lee2023}. Although this problem defies the limits of classical computers, quantum computers offer natural platform to simulate it~\cite{OGorman2022,Liu2023}.
Let us begin by briefly summarizing how to encode the electronic structure problem into a quantum computer~\cite{quantumchemistrytoday,Kassal2011}. The first step is to consider the electronic structure Hamiltonian Hamiltonian $\hat{H}(\boldsymbol{R})=\sum^N_{i}\hat{h}(i)+\sum^N_{i<j}r^{-1}_{i,j}$ under the Born-Oppenheimer approximation~\cite{Born1927}, 
where $N$ is the number of electrons. The term $\hat{h}(i)$ contains information of the kinetic energy and the potential energy of interaction of the $i$-th electron with $M$ nuclei, while $r_{i,j}=|\boldsymbol{r}_{i}-\boldsymbol{r}_{j}|$ is the distance between the $i$-th and $j$-th electrons~\cite{Born1927}. Further, the Hamiltonian depends parametrically on the positions $\boldsymbol{R}=(\boldsymbol{R}_{1}, \boldsymbol{R}_{2}, \dots, \boldsymbol{R}_{M})$ of the $M$ nuclei. 
To account for the fermionic statistics of the electrons, it is convenient to work with the electronic structure Hamiltonian in second quantization
%%%
\begin{align}
        \label{eq:HamiltonianSecondQuantization}
    \hat{H}(\boldsymbol{R})&=\sum_{i,j}h_{i,j}(\boldsymbol{R})\hat{a}^{\dagger}_{i}\hat{a}_j+\frac{1}{2}\sum_{i,j}V_{i,j,k,l}(\boldsymbol{R})\hat{a}^{\dagger}_i\hat{a}^{\dagger}_j\hat{a}_k\hat{a}_l
    \ .
\end{align}
%%%
Here $h_{i,j}(\boldsymbol{R})$ and $V_{i,j,k,l}(\boldsymbol{R})$ denote one- and two-electron integrals that are related to the terms $\hat{h}(i)$ and $1/r_{i,j}$ discussed above. Further, the operators $\hat{a}^{\dagger}_i$ and $\hat{a}_i$ are the fermionic creation and annihilation operators satisfying the anticommutation relations $\{\hat{a}_i, \hat{a}^{\dagger}_j\} =\delta_{i,j}$ and $\{\hat{a}_i, \hat{a}_j\}= \{\hat{a}^{\dagger}_i, \hat{a}^{\dagger}_j\}=0$. 

In our work, we explore how molecular geometry influences operator spreading and entanglement in a one dimensional Hydrogen chain $\text{H}_4$. 
This problem involves $N=4$ electrons and $M=4$ nuclei. The geometry of our system is determined by the nuclear positions $\boldsymbol{R}_{l}=[X_l,0,0]$ for $l=1,\ldots, 4$.  all the coordinates  $X_l=(l-1)R-(-1)^l r$ depends on a single parameter $|r|<3.3a_0$, where $a_0$ is the Bohr radius and  $R=7.55 a_0$. For this reason, from now on, we will abuse the notation and use $\hat{H}(r)$ instead of $\hat{H}(\boldsymbol{R})$ to denote the parametric dependence of the electronic structure Hamiltonian.

Next we use the Jordan-Wigner transformation~\cite{jordan1928} of the fermionic operators $\hat{a}^{\dagger}_j = \left( \bigotimes^{j-1}_{i=1} \hat{\sigma^{z}_i} \right) \otimes \hat{\sigma^{+}_j}$ to represent $\hat{H}(r)$ in terms of Pauli operators for a given parameter $r$ determining the geometry, thus mapping the system to a quantum circuit. We assume that each electron can occupy $4$ delocalized spatial orbitals~\cite{google2020}. That is, if we add the spin degrees of freedom there is a set of $8$ spin orbitals labelled by $j$ in Eq.~\eqref{eq:HamiltonianSecondQuantization}, which allows us to encode the problem  using 8 qubits. We implemented the $H_4$ Hamiltonian in terms of Pauli operators using the \textit{pyscf} library integrated within the \textit{Pennylane} framework \cite{pennylane}. Every Hamiltonian matrix can be decomposed into a sum of Pauli operators and constructed using quantum gates that can be implemented in digital quantum computers \cite{gate_decomposition}.

\textit{Out-of-time-order correlators in electronic structure:}
Now let us evaluate how the molecular geometry influences operator spreading~\cite{LS_echo,operator_spreading,garttner2017} in the qubit representation.
We consider two local unitaries $\hat{V}=\sigma^z_i$ and $\hat{W} = e^{-\mathrm{i}\alpha\hat{\sigma}^z_j/2}$ with $\alpha\in [0, 2\pi]$. As $\sigma^z_i=2\hat{a}^{\dagger}_i\hat{a}_i-1$ in the fermionic representation, $\hat{V}$ gives us the local occupation of $i$-th spin orbital,  while $\hat{W}$ is essentially a local orbital rotation acting on the $j$-th orbital.

From the definition, we can see that $[\hat{V},\hat{W}]=0$. However, if we now consider the operator $\hat{W}$ in the Heisenberg picture $\hat{W}(T)=\hat{U}^{\dagger}(T)\hat{W}\hat{U}(T)$  with $\hat{U}(T)=\exp\left[-\mathrm{i}\hat{H}(r)T\right]$ being the evolution under the electronic structure Hamiltonian, after some time $T$ we will have $[\hat{V},\hat{W}(T)]\neq 0$. 
\begin{figure}[H]
\centering
    \includegraphics[width=0.49 \textwidth]{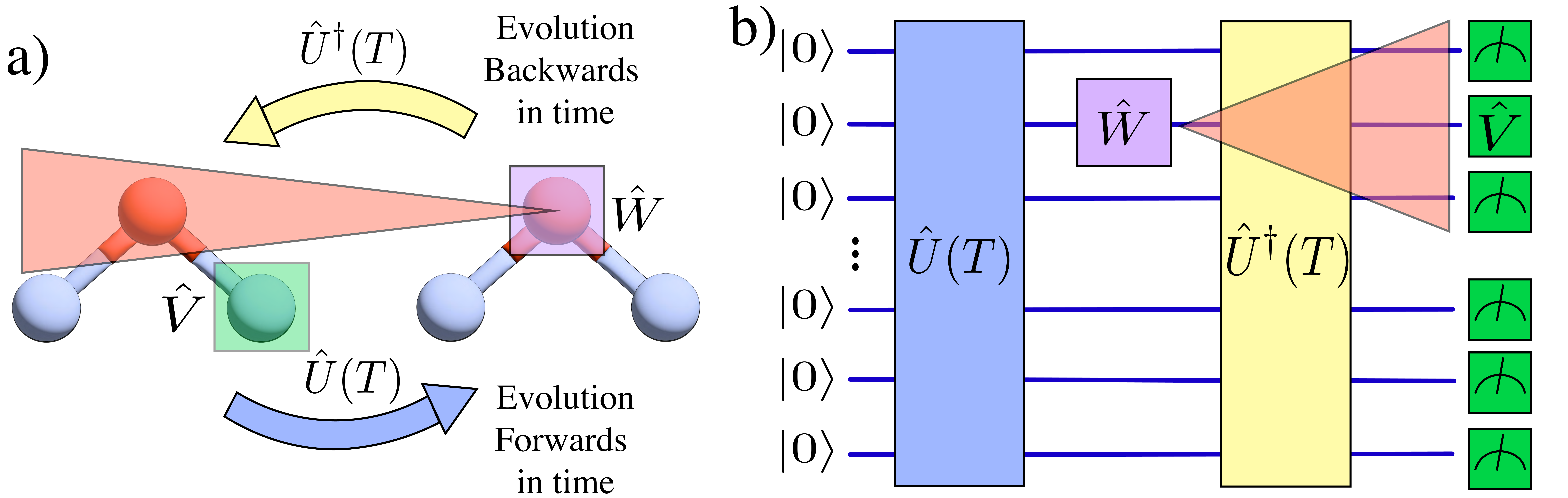}
    \caption{Operator spreading in electronic structure using quantum computers. a) Illustrates  how local operators acting on molecular orbitals propagate during the dynamics. b) Shows the corresponding quantum circuit to implement our idea. The red region represent the light cone characteristic of operator spreading.
    }
    \label{Fig1}
\end{figure}

This is the onset of operator spreading~\cite{Burrell2007} that can captured by the out-of-time order correlation (OTOC)~\cite{LS_echo,operator_spreading,garttner2017}
\begin{align} 
    \label{eq:OperatorSpread}
    \mathcal{F}_{V,W}(T)=\langle\hat{W}^{\dagger}(T)\hat{V}^{\dagger}\hat{W}(T)\hat{V}\rangle
    \ ,
\end{align}
where the expectation value is calculated in the initial state $\ket{\psi_{\text{in}}(0)}$. This gives us information about operator spreading because $\Re[\mathcal{F}_{V,W}(T)]=1-\langle|[\hat{V},\hat{W}(T)]|^2\rangle/2$~\cite{garttner2017,Shen2020}. 
Note that the operators $\hat{V}$ and $\hat{W}$ we consider in this work are local in the fermionic and qubit representations and the concept of operator spreading of local operators makes sense from both points of view. This is not the case for other operators such as $\hat{a}^{\dagger}_j$, because they are local in the fermionic representation but highly nonlocal in terms of spins.

Figs.~\ref{Fig1}~a)~and~b) illustrate an operational way to measure the OTOCs in a molecular system by applying local unitaries and time reversal in a quantum computer~\cite{Shen2020}. To calculate $\mathcal{F}_{V,W}(T)$ and for the all the calculations in this work, we use initial state $\ket{\psi_{\text{in}}(0)}= |11110000 \rangle$ that is easy to prepare in a quantum computer. For this initial state the OTOC can be rewritten as $\mathcal{F}_{V,W}(T)=\pm\langle \psi_{\text{fin}}(T)|\sigma^z_i|\psi_{\text{fin}}(T)\rangle$, where $\ket{\psi_{\text{fin}}(T)}=\hat{W}(T)\ket{\psi_{\text{in}}(0)}$ can be prepared using the quantum circuit in Fig.~\ref{Fig1}~b). Note that the $\pm$ sign in our OTOC depends on the orbital label $i$ at which we perform the measurement.
To prepare $|\psi_{\text{fin}}(T)\rangle$ can be quite challenging as it involves to perform evolution backwards in time, but this can be performed in a quantum computer~\cite{Shen2020}.

\begin{figure}[H]
        \centering
        \includegraphics[width=0.5\textwidth]{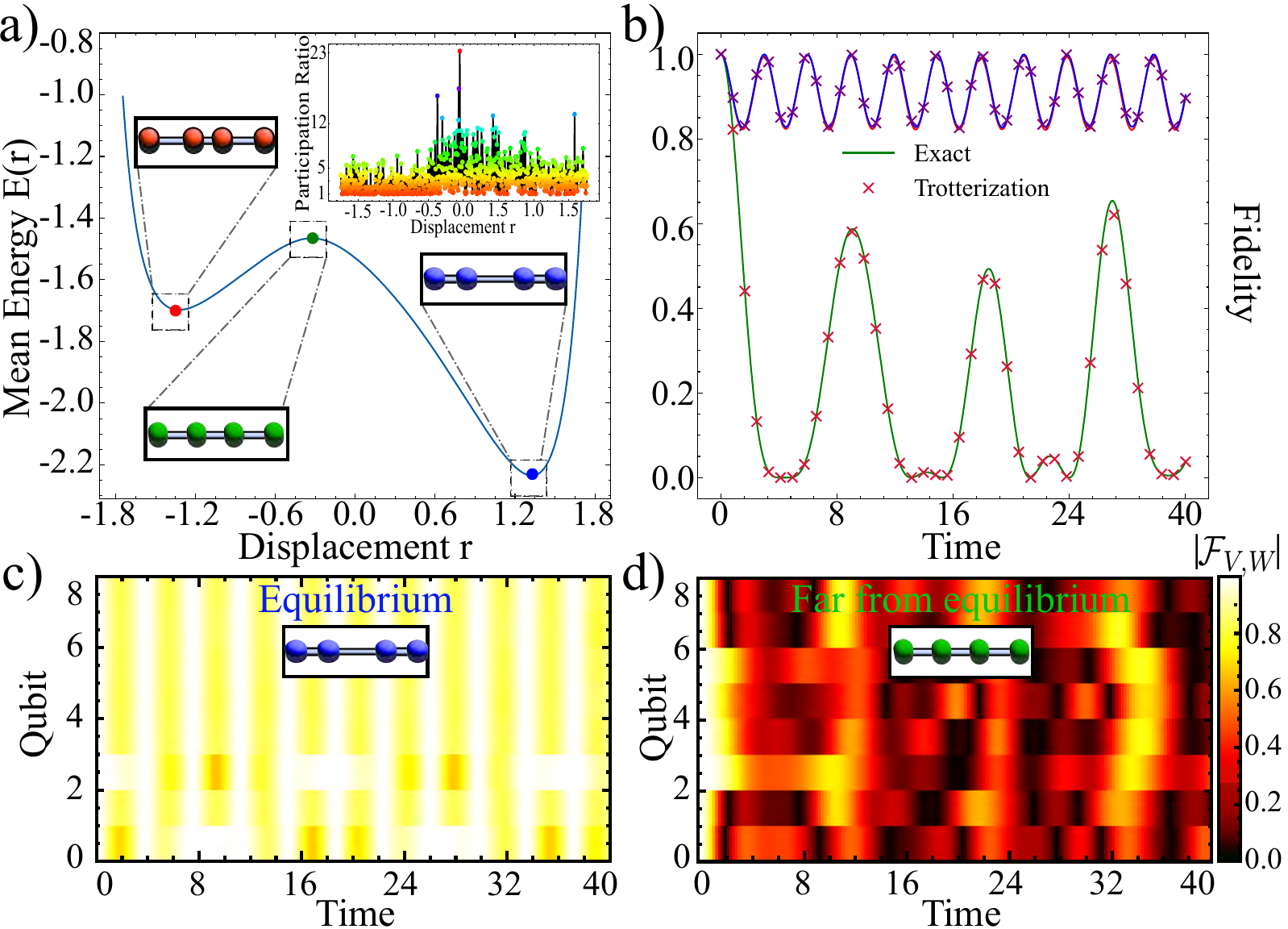}
        \caption{Molecular geometry, fidelity and OTOCs. a) Depicts the mean energy landscape as a function of $r$. The insets illustrate three different geometries of our hydrogen chain. 
        In b) we show the fidelity  
        for these geometries, comparing results obtained from both exact and Trotterized simulations and by applying a local orbital rotation $\hat{W} = e^{-\mathrm{i}\alpha\hat{\sigma}^z_1/2}$ to the first qubit with $\alpha=\pi$. Figures c)~and~d) Show the dynamics for our OTOC $\mathcal{F}_{V,W}(T)$ for $\hat{V}=\sigma^z_i$ for different orbitals labelled by $i$.
        }
         \label{Fig2}
        \end{figure}

\textit{Mean energy surfaces and operator spreading:}
It is important to explore in more detail how molecular geometry influences the quantum dynamics. Previously, we consider an initial $\ket{\psi_{\text{in}}(0)}= |11110000 \rangle$ that is not the ground state nor any of the excited states in electronic structure, which are highly entangled~\cite{boguslawski2013orbital}. Our initial state is a linear combination $\ket{\psi_{\text{in}}(0)}=\sum_n c_n(r)\ket{E_n(r)}$ of molecular energy eigenstates that are solutions of the electronic structure problem $\hat{H}(r)\ket{E_n(r)}=E_n(r)\ket{E_n(r)}$. The parametric dependence indicates that they are determined by the geometry of the molecule thus defining a family $\{E_n(r)\}$ of energy surfaces~\cite{hoffmann1970geometry}. In other words, given $E_n(r)$, there is an associated geometry of the molecule~\cite{hoffmann1970geometry}. For example,  for hydrogen cyanide HCN, the ground state energy exhibits a linear equilibrium geometry while its three excited single states are bent~\cite{hoffmann1970geometry}. Motivated by this, in our work, we define the mean energy landscape associated to the initial state, as follows

\begin{align}
     \bar{E}(r)=\text{Tr}[\hat{H}(r)\rho_{\text{in}}]=\sum_n |c_n(r)|^2E_n(r)
     .
   \label{eq:MeanEnergyLandscape}
\end{align}
From this one can see that $|c_n(r)|^2$ give us information of how much a given energy surface contribute to $\bar{E}(r)$.
Figure~\ref{Fig2}~a) shows the energy landscape for our hydrogen chain as a function of the parameter $r$. This diagram clearly shows two low-energy equilibrium geometries (red and blue) and an unstable transition state (green). It is important to have a chemical intuition of these geometries. When two hydrogen atoms approach, they form bonding and antibonding molecular orbitals. The lowest energy configuration corresponds then to the bonding orbital. For this reason, the lowest energy is achieved in our hydrogen chain when the atoms form two dimers (blue). Bringing the atoms far away from the equilibrium geometry implies to have a larger energy (green).

To quantify how the geometry of the surface influences the localization of our initial state $\ket{\psi_{\text{in}}(0)}$ in the energy eigenbasis, we consider 
the participation ratio $P^{-1}(r) = 1/\left(\sum |c_n(r)|^4\right)$ for different values of $r$. The inset of Fig.~\ref{Fig2}~a) shows that the participation ratio at the unstable geometry is large in comparison to the equilibrium points. 

Next, we will investigate how the mean energy surface associated to the initial state and the participation ratio $P^{-1}(r)$ determine the nature of operator spreading. As we discussed in the previous section, our protocol to calculate $\mathcal{F}_{V,W}(T)$ involves a local orbital rotation $\hat{W}$ that propagates backwards in time and spreads out across the system\cite{operator_spreading}. To quantify this, we calculate the fidelity $F_{r}(T) = |\langle \psi_{\text{in}}(0)|\psi_{\text{fin}}(T)\rangle|^2$ between the initial state and the final state, which allows to assess the system's sensitivity to small local orbital rotations for different geometries determined by $\bar{E}(r)$ in Fig.~\ref{Fig2}~a). 

Before discussing our numerical results for the fidelity and the OTOC it is important to understand how these quantities are related to the energy surfaces $\{E_m(r)\}$. Let us start by considering the operator in the energy eigenbasis 
\begin{align}
     \hat{W}(T)=\sum_{m,n}e^{\mathrm{i}\omega_{m,n}T}W_{m,n}\ket{E_m(r)}\bra{E_n(r)}
     \ ,
   \label{eq:SurfaceHoppingOperator}
\end{align}
where
$W_{m,n}=\bra{E_m(r)}\hat{W}\ket{E_n(r)}$ and $\omega_{m,n}=E_m(r)-E_n(r)$. This simple expression together with the mean energy landscape and the participation ratio, is all what we need to know to understand the dynamics of the fidelity and the OTOC.

Let us first start by considering the transition probability $\langle\psi_{\text{in}}(0)\ket{\psi_{\text{fin}}(T)}=\langle\psi_{\text{in}}(0)|\hat{W}(T)|\psi_{\text{in}}(0)\rangle$. Given the initial state $\ket{\psi_{\text{in}}(0)}=\sum_n c_n(r)\ket{E_n(r)}$, $\hat{W}(T)$ induces electronic transitions or "surface hoppings" that are weighted by the coefficients $c_n$. 
At the equilibrium geometries, the participation ratio $P^{-1}(r)$ is low and our state is localized in the energy eigenbasis. 
For this reason the transitions induced by $\hat{W}$ are restricted and the few oscillation frequencies $\omega_{m,n}$ in Eq.~\eqref{eq:SurfaceHoppingOperator} give us a periodic behavior of fidelity as in Fig.~\ref{Fig2} b). 

Contrary to this, when we are far from the equilibrium geometry, the initial state is a superposition of many eigenstates with larger participation ratio. Thus, there are more transitions inducing a decay of the fidelity and the oscillations become non-harmonic because there are more relevant frequencies, as we show in Fig.~\ref{Fig2} b). 
 
We can use similar arguments to understand the OTOC
\begin{align}
     \mathcal{F}_{V,W}(T)=\sum_{m,n,k,l}e^{\mathrm{i}(\omega_{m,n}-\omega_{k,l})T}W_{m,n}W^*_{k,l}V_{l,m}c_k^*c_n
     \ ,
   \label{eq:SurfaceHoppingOperator}
\end{align}
where $V_{l,m}=\bra{E_m(r)}\hat{V}\ket{E_n(r)}$. Thus, the OTOC can be interpreted as sequence of "surface hoppings" generated by $\hat{W}$ and $\hat{V}$ and weighted by the coefficients $c_n$. Figs.~\ref{Fig2}~c)~and~d) depict the dynamics of  $|\mathcal{F}_{V,W}(T)|$. Clearly, our result is consistent with the fidelity in Fig.~\ref{Fig2}~b) and reveals that the system is extremely sensible to the action of local unitaries when it is far from the equilibrium geometry and exhibits a stronger operator spreading with a non-periodic behavior in time. The reason for this is the large amount of energy surfaces available to hop to in this configuration.

\textit{Molecular geometry and entanglement entropy:}
Operator spreading and OTOCs are related to fascinating concepts in manybody and statistical physics with applications in diverse fields~\cite{LS_echo,operator_spreading,garttner2017}. In the manybody localized phase, local conserved quantities restrict the operator spreading and the bipartite entanglement exhibits an area law~\cite{bauer2013area,area_law}. On the contrary, ergodic system have large operator spreading and the bipartite entanglement shows a volume law~\cite{Bertini2019}.

It is natural to think that if we observe operator spreading in electronic structure, this will be accompanied of entanglement growth between in the system. As the molecular orbitals are arranged in a one-dimensional array so we can use the Jordan-Wigner transformation, after qubit mapping our system is encoded in a quantum computer using a one-dimensional spin chain with $8$ qubits and nonlocal interactions. Motivated by this, we will calculate a measure of bipartite entanglement and study its dependence with the length of a bipartition of the chain into two parts $A$ and $B$ with $A$ having $L=2,3,4$ qubits.
More specifically, we consider the state $\ket{\psi(T)}=\hat{U}(T)\ket{\psi_{\text{in}}(0)}$ with $\hat{U}(T)=\exp\left[-\mathrm{i}\hat{H}(r)T\right]$ for a fixed geometry $r$. As a next step, we define the reduced density matrix $\hat{\rho}_L(t) = \text{Tr}_{B}[\hat{\rho}(T)]$ of the density matrix $\hat{\rho}(T)=\ket{\psi(T)}\bra{\psi(T)}$ and use the von-Neumann entropy~\cite{area_law} defined as $S_L=-\text{Tr}_A[\hat{\rho}_L \log (\hat{\rho}_L)]$. It is important to remind the reader on the basics of entanglement entropy and the limiting cases. For a separable quantum state $\ket{\psi(T)}$ of the two subsystems $A$ and $B$ discussed above, the reduced density matrix $\hat{\rho}_L(t)$ turns out to be pure. In this case,   $S(\hat{\rho}_L) = 0$. In other situations the value is going to increase dependent on how mixed the reduced density matrix $\hat{\rho}_L$ is. 

Entanglement also has implications in quantum chemistry~\cite{molina2015quantum,esquivel2015quantum} ans it intimately related to the correlation energy~\cite{boguslawski2013orbital,ding2020concept}. For this reason, it is interesting to explore the dependence of entanglement entropy of our molecular system encoded as a one dimensional spin chain as a function of the partition size $L$. In systems that satisfy an area law, the entropy of the system increases in proportion to its surface area rather than its volume. This means that regardless of the number of qubits involved in the bipartition, the entropy does not necessarily increase directly with the partition's size $L$. In contrast, for systems with volume law, the entanglement entropy scales with the subsystem's size $L$.

\begin{figure}[H]
    \centering
    \includegraphics[width=0.48 \textwidth]{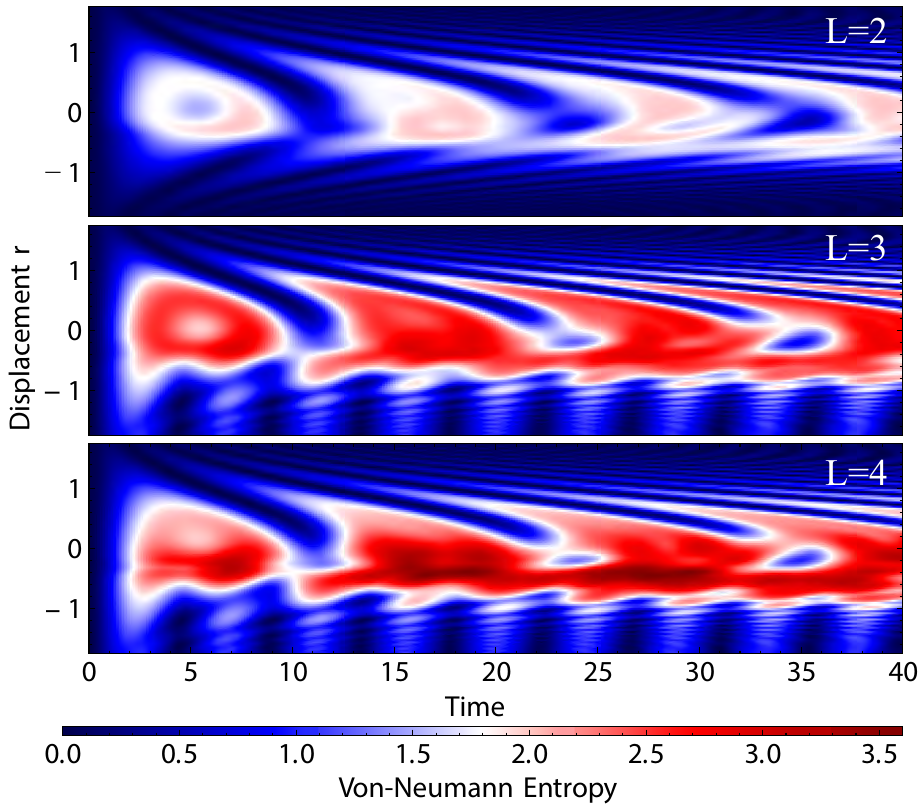}
    \caption{Entanglement entropy in the Hydrogen chain $H_4$ for different lengths $L=2,3 \ \text{and} \ 4$  of the bipartition A and B of the system. 
    The upper, middle and lower panels show the entanglement entropy for the reduced density matrices $\hat{\rho}_2$, $\hat{\rho}_3$, and $\hat{\rho}_4$, respectively. Our results reveal the entanglement grows with the subsystem size $L$ when the system is far from the equilibrium geometry.
    }
    \label{Fig3}
\end{figure}

To have insight on the size dependence of the entanglement entropy for our system, we calculate its dependence for different values of $r$ and $L$. The results are depicted in Fig.~\ref{Fig3}. Our calculations show that a clear trend emerges regarding the entanglement entropy within the system. Moving from $\rho_{2}$, $\rho_{3}$ to $\rho_{4}$, we observe increasing entropy values when the system is far from equilibrium geometry ($r\approx 0$). This behavior suggests a correlation between the size $L$ of the region under consideration and the entanglement entropy, which is a pattern that is expected for systems under a volume law. Contrary to this, when the system is close to equilibrium geometries the values of entanglement entropy for the three subsystems $\rho_{1}$, $\rho_{2}$, and $\rho_{3}$  do not increase with system size and show with a time-periodic behaviour consistent with the periodicity of the fidelity from figure \ref{Fig2}~b).
This result resembles an area law as the entanglement is not sensible to the length $L$ of the partition.

Our analysis of entanglement for different partition sizes $L$ indicates a potential relation the entanglement entropy with the partition size $L$ and the molecular geometry of the system. 
This naturally offers valuable insights into the nature of entanglement when the system is equilibrium and far-from-equilibrium geometries, respectively. Our results have implications for quantum simulation of electronic structure using quantum computers as information scrambling implies that the information becomes non-local during the time evolution for far-from-equilibrium geometries.
In a quantum simulation, local coherent errors can be interpreted as local operators $\hat{W}$, which at short times are local, but that rapidly spread out in the syste and become correlated.

\textit{Conclusions:}  
In summary, we have defined out-of-time order correlators for electronic structure problems encoded in a quantum computer. 
We show that the mean energy surface and the associated molecular geometries strongly influence the operator spreading and the behavior of the bipartite entanglement. 
Of course, we are not restricted to highly symmetric molecules such as our hydrogen chain.
We could, for instance  consider other molecules with less symmetries and investigate if area-like and volume-like laws are generic in molecular systems when they are in equilibrium and far-from equilibrium geometries, respectively. Other direction of research is to investigate the implications of our results on
error propagation of coherent errors due to hardware in quantum simulations of electronic structure.

{\it{Acknowledgments.---} }
%%%%%%%%%%%%%%%%%%%%%
The authors thank M. Cho, A. Gangat, T. Van Voorhis, J. Riedel, H. Sato, and S. Weatherly for valuable discussions. The authors acknowledge NTT Research Inc.  for their support during this project.


\begin{thebibliography}{52}%
\makeatletter
\providecommand \@ifxundefined [1]{%
 \@ifx{#1\undefined}
}%
\providecommand \@ifnum [1]{%
 \ifnum #1\expandafter \@firstoftwo
 \else \expandafter \@secondoftwo
 \fi
}%
\providecommand \@ifx [1]{%
 \ifx #1\expandafter \@firstoftwo
 \else \expandafter \@secondoftwo
 \fi
}%
\providecommand \natexlab [1]{#1}%
\providecommand \enquote  [1]{``#1''}%
\providecommand \bibnamefont  [1]{#1}%
\providecommand \bibfnamefont [1]{#1}%
\providecommand \citenamefont [1]{#1}%
\providecommand \href@noop [0]{\@secondoftwo}%
\providecommand \href [0]{\begingroup \@sanitize@url \@href}%
\providecommand \@href[1]{\@@startlink{#1}\@@href}%
\providecommand \@@href[1]{\endgroup#1\@@endlink}%
\providecommand \@sanitize@url [0]{\catcode `\\12\catcode `\$12\catcode
  `\&12\catcode `\#12\catcode `\^12\catcode `\_12\catcode `\%12\relax}%
\providecommand \@@startlink[1]{}%
\providecommand \@@endlink[0]{}%
\providecommand \url  [0]{\begingroup\@sanitize@url \@url }%
\providecommand \@url [1]{\endgroup\@href {#1}{\urlprefix }}%
\providecommand \urlprefix  [0]{URL }%
\providecommand \Eprint [0]{\href }%
\providecommand \doibase [0]{http://dx.doi.org/}%
\providecommand \selectlanguage [0]{\@gobble}%
\providecommand \bibinfo  [0]{\@secondoftwo}%
\providecommand \bibfield  [0]{\@secondoftwo}%
\providecommand \translation [1]{[#1]}%
\providecommand \BibitemOpen [0]{}%
\providecommand \bibitemStop [0]{}%
\providecommand \bibitemNoStop [0]{.\EOS\space}%
\providecommand \EOS [0]{\spacefactor3000\relax}%
\providecommand \BibitemShut  [1]{\csname bibitem#1\endcsname}%
\let\auto@bib@innerbib\@empty
%</preamble>
\bibitem [{\citenamefont {Kohn}(1999)}]{Kohn1999}%
  \BibitemOpen
  \bibfield  {author} {\bibinfo {author} {\bibfnamefont {W.}~\bibnamefont
  {Kohn}},\ }\href {\doibase 10.1103/RevModPhys.71.1253} {\bibfield  {journal}
  {\bibinfo  {journal} {Rev. Mod. Phys.}\ }\textbf {\bibinfo {volume} {71}},\
  \bibinfo {pages} {1253} (\bibinfo {year} {1999})}\BibitemShut {NoStop}%
\bibitem [{\citenamefont {Whitfield}\ \emph {et~al.}(2013)\citenamefont
  {Whitfield}, \citenamefont {Love},\ and\ \citenamefont
  {Aspuru-Guzik}}]{whitfield2013}%
  \BibitemOpen
  \bibfield  {author} {\bibinfo {author} {\bibfnamefont {J.~D.}\ \bibnamefont
  {Whitfield}}, \bibinfo {author} {\bibfnamefont {P.~J.}\ \bibnamefont {Love}},
  \ and\ \bibinfo {author} {\bibfnamefont {A.}~\bibnamefont {Aspuru-Guzik}},\
  }\href@noop {} {\bibfield  {journal} {\bibinfo  {journal} {Physical Chemistry
  Chemical Physics}\ }\textbf {\bibinfo {volume} {15}},\ \bibinfo {pages} {397}
  (\bibinfo {year} {2013})}\BibitemShut {NoStop}%
\bibitem [{\citenamefont {McArdle}\ \emph {et~al.}(2020)\citenamefont
  {McArdle}, \citenamefont {Endo}, \citenamefont {Aspuru-Guzik}, \citenamefont
  {Benjamin},\ and\ \citenamefont {Yuan}}]{McArdle2020}%
  \BibitemOpen
  \bibfield  {author} {\bibinfo {author} {\bibfnamefont {S.}~\bibnamefont
  {McArdle}}, \bibinfo {author} {\bibfnamefont {S.}~\bibnamefont {Endo}},
  \bibinfo {author} {\bibfnamefont {A.}~\bibnamefont {Aspuru-Guzik}}, \bibinfo
  {author} {\bibfnamefont {S.~C.}\ \bibnamefont {Benjamin}}, \ and\ \bibinfo
  {author} {\bibfnamefont {X.}~\bibnamefont {Yuan}},\ }\href {\doibase
  10.1103/RevModPhys.92.015003} {\bibfield  {journal} {\bibinfo  {journal}
  {Rev. Mod. Phys.}\ }\textbf {\bibinfo {volume} {92}},\ \bibinfo {pages}
  {015003} (\bibinfo {year} {2020})}\BibitemShut {NoStop}%
\bibitem [{\citenamefont {Deumens}\ \emph {et~al.}(1994)\citenamefont
  {Deumens}, \citenamefont {Diz}, \citenamefont {Longo},\ and\ \citenamefont
  {\"Ohrn}}]{Deumens1994}%
  \BibitemOpen
  \bibfield  {author} {\bibinfo {author} {\bibfnamefont {E.}~\bibnamefont
  {Deumens}}, \bibinfo {author} {\bibfnamefont {A.}~\bibnamefont {Diz}},
  \bibinfo {author} {\bibfnamefont {R.}~\bibnamefont {Longo}}, \ and\ \bibinfo
  {author} {\bibfnamefont {Y.}~\bibnamefont {\"Ohrn}},\ }\href {\doibase
  10.1103/RevModPhys.66.917} {\bibfield  {journal} {\bibinfo  {journal} {Rev.
  Mod. Phys.}\ }\textbf {\bibinfo {volume} {66}},\ \bibinfo {pages} {917}
  (\bibinfo {year} {1994})}\BibitemShut {NoStop}%
\bibitem [{\citenamefont {Lisinetskaya}\ and\ \citenamefont
  {Mitri\ifmmode~\acute{c}\else \'{c}\fi{}}(2011)}]{Lisinetskaya2011}%
  \BibitemOpen
  \bibfield  {author} {\bibinfo {author} {\bibfnamefont {P.~G.}\ \bibnamefont
  {Lisinetskaya}}\ and\ \bibinfo {author} {\bibfnamefont {R.}~\bibnamefont
  {Mitri\ifmmode~\acute{c}\else \'{c}\fi{}}},\ }\href {\doibase
  10.1103/PhysRevA.83.033408} {\bibfield  {journal} {\bibinfo  {journal} {Phys.
  Rev. A}\ }\textbf {\bibinfo {volume} {83}},\ \bibinfo {pages} {033408}
  (\bibinfo {year} {2011})}\BibitemShut {NoStop}%
\bibitem [{\citenamefont {Born}\ and\ \citenamefont
  {Oppenheimer}(1927)}]{Born1927}%
  \BibitemOpen
  \bibfield  {author} {\bibinfo {author} {\bibfnamefont {M.}~\bibnamefont
  {Born}}\ and\ \bibinfo {author} {\bibfnamefont {R.}~\bibnamefont
  {Oppenheimer}},\ }\href {\doibase https://doi.org/10.1002/andp.19273892002}
  {\bibfield  {journal} {\bibinfo  {journal} {Annalen der Physik}\ }\textbf
  {\bibinfo {volume} {389}},\ \bibinfo {pages} {457} (\bibinfo {year}
  {1927})}\BibitemShut {NoStop}%
\bibitem [{\citenamefont {O'Gorman}\ \emph {et~al.}(2022)\citenamefont
  {O'Gorman}, \citenamefont {Irani}, \citenamefont {Whitfield},\ and\
  \citenamefont {Fefferman}}]{OGorman2022}%
  \BibitemOpen
  \bibfield  {author} {\bibinfo {author} {\bibfnamefont {B.}~\bibnamefont
  {O'Gorman}}, \bibinfo {author} {\bibfnamefont {S.}~\bibnamefont {Irani}},
  \bibinfo {author} {\bibfnamefont {J.}~\bibnamefont {Whitfield}}, \ and\
  \bibinfo {author} {\bibfnamefont {B.}~\bibnamefont {Fefferman}},\ }\href
  {\doibase 10.1103/PRXQuantum.3.020322} {\bibfield  {journal} {\bibinfo
  {journal} {PRX Quantum}\ }\textbf {\bibinfo {volume} {3}},\ \bibinfo {pages}
  {020322} (\bibinfo {year} {2022})}\BibitemShut {NoStop}%
\bibitem [{\citenamefont {Cao}\ \emph {et~al.}(2019)\citenamefont {Cao},
  \citenamefont {Romero}, \citenamefont {Olson}, \citenamefont {Degroote},
  \citenamefont {Johnson}, \citenamefont {Kieferová}, \citenamefont
  {Kivlichan}, \citenamefont {Menke}, \citenamefont {Peropadre}, \citenamefont
  {Sawaya}, \citenamefont {Sim}, \citenamefont {Veis},\ and\ \citenamefont
  {Aspuru-Guzik}}]{quantumchemistrytoday}%
  \BibitemOpen
  \bibfield  {author} {\bibinfo {author} {\bibfnamefont {Y.}~\bibnamefont
  {Cao}}, \bibinfo {author} {\bibfnamefont {J.}~\bibnamefont {Romero}},
  \bibinfo {author} {\bibfnamefont {J.~P.}\ \bibnamefont {Olson}}, \bibinfo
  {author} {\bibfnamefont {M.}~\bibnamefont {Degroote}}, \bibinfo {author}
  {\bibfnamefont {P.~D.}\ \bibnamefont {Johnson}}, \bibinfo {author}
  {\bibfnamefont {M.}~\bibnamefont {Kieferová}}, \bibinfo {author}
  {\bibfnamefont {I.~D.}\ \bibnamefont {Kivlichan}}, \bibinfo {author}
  {\bibfnamefont {T.}~\bibnamefont {Menke}}, \bibinfo {author} {\bibfnamefont
  {B.}~\bibnamefont {Peropadre}}, \bibinfo {author} {\bibfnamefont {N.~P.~D.}\
  \bibnamefont {Sawaya}}, \bibinfo {author} {\bibfnamefont {S.}~\bibnamefont
  {Sim}}, \bibinfo {author} {\bibfnamefont {L.}~\bibnamefont {Veis}}, \ and\
  \bibinfo {author} {\bibfnamefont {A.}~\bibnamefont {Aspuru-Guzik}},\ }\href
  {\doibase 10.1021/acs.chemrev.8b00803} {\bibfield  {journal} {\bibinfo
  {journal} {Chemical Reviews}\ }\textbf {\bibinfo {volume} {119}},\ \bibinfo
  {pages} {10856} (\bibinfo {year} {2019})}\BibitemShut {NoStop}%
\bibitem [{\citenamefont {Gujarati}\ \emph {et~al.}(2023)\citenamefont
  {Gujarati}, \citenamefont {Motta}, \citenamefont {Friedhoff} \emph
  {et~al.}}]{perturbations}%
  \BibitemOpen
  \bibfield  {author} {\bibinfo {author} {\bibfnamefont {T.~P.}\ \bibnamefont
  {Gujarati}}, \bibinfo {author} {\bibfnamefont {M.}~\bibnamefont {Motta}},
  \bibinfo {author} {\bibfnamefont {T.~N.}\ \bibnamefont {Friedhoff}},  \emph
  {et~al.},\ }\href {\doibase 10.1038/s41534-023-00753-1} {\bibfield  {journal}
  {\bibinfo  {journal} {npj Quantum Inf}\ }\textbf {\bibinfo {volume} {9}},\
  \bibinfo {pages} {88} (\bibinfo {year} {2023})}\BibitemShut {NoStop}%
\bibitem [{\citenamefont {Kassal}\ \emph {et~al.}(2011)\citenamefont {Kassal},
  \citenamefont {Whitfield}, \citenamefont {Perdomo-Ortiz}, \citenamefont
  {Yung},\ and\ \citenamefont {Aspuru-Guzik}}]{Kassal2011}%
  \BibitemOpen
  \bibfield  {author} {\bibinfo {author} {\bibfnamefont {I.}~\bibnamefont
  {Kassal}}, \bibinfo {author} {\bibfnamefont {J.~D.}\ \bibnamefont
  {Whitfield}}, \bibinfo {author} {\bibfnamefont {A.}~\bibnamefont
  {Perdomo-Ortiz}}, \bibinfo {author} {\bibfnamefont {M.-H.}\ \bibnamefont
  {Yung}}, \ and\ \bibinfo {author} {\bibfnamefont {A.}~\bibnamefont
  {Aspuru-Guzik}},\ }\href {\doibase
  https://doi.org/10.1146/annurev-physchem-032210-103512} {\bibfield  {journal}
  {\bibinfo  {journal} {Annual Review of Physical Chemistry}\ }\textbf
  {\bibinfo {volume} {62}},\ \bibinfo {pages} {185} (\bibinfo {year}
  {2011})}\BibitemShut {NoStop}%
\bibitem [{\citenamefont {Bharti}\ \emph {et~al.}(2022)\citenamefont {Bharti},
  \citenamefont {Cervera-Lierta}, \citenamefont {Kyaw}, \citenamefont {Haug},
  \citenamefont {Alperin-Lea}, \citenamefont {Anand}, \citenamefont {Degroote},
  \citenamefont {Heimonen}, \citenamefont {Kottmann}, \citenamefont {Menke},
  \citenamefont {Mok}, \citenamefont {Sim}, \citenamefont {Kwek},\ and\
  \citenamefont {Aspuru-Guzik}}]{Bharti2022}%
  \BibitemOpen
  \bibfield  {author} {\bibinfo {author} {\bibfnamefont {K.}~\bibnamefont
  {Bharti}}, \bibinfo {author} {\bibfnamefont {A.}~\bibnamefont
  {Cervera-Lierta}}, \bibinfo {author} {\bibfnamefont {T.~H.}\ \bibnamefont
  {Kyaw}}, \bibinfo {author} {\bibfnamefont {T.}~\bibnamefont {Haug}}, \bibinfo
  {author} {\bibfnamefont {S.}~\bibnamefont {Alperin-Lea}}, \bibinfo {author}
  {\bibfnamefont {A.}~\bibnamefont {Anand}}, \bibinfo {author} {\bibfnamefont
  {M.}~\bibnamefont {Degroote}}, \bibinfo {author} {\bibfnamefont
  {H.}~\bibnamefont {Heimonen}}, \bibinfo {author} {\bibfnamefont {J.~S.}\
  \bibnamefont {Kottmann}}, \bibinfo {author} {\bibfnamefont {T.}~\bibnamefont
  {Menke}}, \bibinfo {author} {\bibfnamefont {W.-K.}\ \bibnamefont {Mok}},
  \bibinfo {author} {\bibfnamefont {S.}~\bibnamefont {Sim}}, \bibinfo {author}
  {\bibfnamefont {L.-C.}\ \bibnamefont {Kwek}}, \ and\ \bibinfo {author}
  {\bibfnamefont {A.}~\bibnamefont {Aspuru-Guzik}},\ }\href {\doibase
  10.1103/RevModPhys.94.015004} {\bibfield  {journal} {\bibinfo  {journal}
  {Rev. Mod. Phys.}\ }\textbf {\bibinfo {volume} {94}},\ \bibinfo {pages}
  {015004} (\bibinfo {year} {2022})}\BibitemShut {NoStop}%
\bibitem [{\citenamefont {Kim}\ \emph {et~al.}(2023)\citenamefont {Kim},
  \citenamefont {Eddins}, \citenamefont {Anand}, \citenamefont {Wei},
  \citenamefont {Van Den~Berg}, \citenamefont {Rosenblatt}, \citenamefont
  {Nayfeh}, \citenamefont {Wu}, \citenamefont {Zaletel}, \citenamefont {Temme}
  \emph {et~al.}}]{Kim2023}%
  \BibitemOpen
  \bibfield  {author} {\bibinfo {author} {\bibfnamefont {Y.}~\bibnamefont
  {Kim}}, \bibinfo {author} {\bibfnamefont {A.}~\bibnamefont {Eddins}},
  \bibinfo {author} {\bibfnamefont {S.}~\bibnamefont {Anand}}, \bibinfo
  {author} {\bibfnamefont {K.~X.}\ \bibnamefont {Wei}}, \bibinfo {author}
  {\bibfnamefont {E.}~\bibnamefont {Van Den~Berg}}, \bibinfo {author}
  {\bibfnamefont {S.}~\bibnamefont {Rosenblatt}}, \bibinfo {author}
  {\bibfnamefont {H.}~\bibnamefont {Nayfeh}}, \bibinfo {author} {\bibfnamefont
  {Y.}~\bibnamefont {Wu}}, \bibinfo {author} {\bibfnamefont {M.}~\bibnamefont
  {Zaletel}}, \bibinfo {author} {\bibfnamefont {K.}~\bibnamefont {Temme}},
  \emph {et~al.},\ }\href@noop {} {\bibfield  {journal} {\bibinfo  {journal}
  {Nature}\ }\textbf {\bibinfo {volume} {618}},\ \bibinfo {pages} {500}
  (\bibinfo {year} {2023})}\BibitemShut {NoStop}%
\bibitem [{\citenamefont {Huh}\ \emph {et~al.}(2015)\citenamefont {Huh},
  \citenamefont {Guerreschi}, \citenamefont {Peropadre}, \citenamefont
  {McClean},\ and\ \citenamefont {Aspuru-Guzik}}]{huh2015}%
  \BibitemOpen
  \bibfield  {author} {\bibinfo {author} {\bibfnamefont {J.}~\bibnamefont
  {Huh}}, \bibinfo {author} {\bibfnamefont {G.~G.}\ \bibnamefont {Guerreschi}},
  \bibinfo {author} {\bibfnamefont {B.}~\bibnamefont {Peropadre}}, \bibinfo
  {author} {\bibfnamefont {J.~R.}\ \bibnamefont {McClean}}, \ and\ \bibinfo
  {author} {\bibfnamefont {A.}~\bibnamefont {Aspuru-Guzik}},\ }\href@noop {}
  {\bibfield  {journal} {\bibinfo  {journal} {Nature Photonics}\ }\textbf
  {\bibinfo {volume} {9}},\ \bibinfo {pages} {615} (\bibinfo {year}
  {2015})}\BibitemShut {NoStop}%
\bibitem [{\citenamefont {Kandala}\ \emph {et~al.}(2017)\citenamefont
  {Kandala}, \citenamefont {Mezzacapo}, \citenamefont {Temme}, \citenamefont
  {Takita}, \citenamefont {Brink}, \citenamefont {Chow},\ and\ \citenamefont
  {Gambetta}}]{kandala2017}%
  \BibitemOpen
  \bibfield  {author} {\bibinfo {author} {\bibfnamefont {A.}~\bibnamefont
  {Kandala}}, \bibinfo {author} {\bibfnamefont {A.}~\bibnamefont {Mezzacapo}},
  \bibinfo {author} {\bibfnamefont {K.}~\bibnamefont {Temme}}, \bibinfo
  {author} {\bibfnamefont {M.}~\bibnamefont {Takita}}, \bibinfo {author}
  {\bibfnamefont {M.}~\bibnamefont {Brink}}, \bibinfo {author} {\bibfnamefont
  {J.~M.}\ \bibnamefont {Chow}}, \ and\ \bibinfo {author} {\bibfnamefont
  {J.~M.}\ \bibnamefont {Gambetta}},\ }\href@noop {} {\bibfield  {journal}
  {\bibinfo  {journal} {nature}\ }\textbf {\bibinfo {volume} {549}},\ \bibinfo
  {pages} {242} (\bibinfo {year} {2017})}\BibitemShut {NoStop}%
\bibitem [{\citenamefont {Kandala}\ \emph {et~al.}(2018)\citenamefont
  {Kandala}, \citenamefont {Temme}, \citenamefont {Corcoles}, \citenamefont
  {Mezzacapo}, \citenamefont {Chow},\ and\ \citenamefont
  {Gambetta}}]{kandala2018}%
  \BibitemOpen
  \bibfield  {author} {\bibinfo {author} {\bibfnamefont {A.}~\bibnamefont
  {Kandala}}, \bibinfo {author} {\bibfnamefont {K.}~\bibnamefont {Temme}},
  \bibinfo {author} {\bibfnamefont {A.~D.}\ \bibnamefont {Corcoles}}, \bibinfo
  {author} {\bibfnamefont {A.}~\bibnamefont {Mezzacapo}}, \bibinfo {author}
  {\bibfnamefont {J.~M.}\ \bibnamefont {Chow}}, \ and\ \bibinfo {author}
  {\bibfnamefont {J.~M.}\ \bibnamefont {Gambetta}},\ }\href@noop {} {\bibfield
  {journal} {\bibinfo  {journal} {arXiv preprint arXiv:1805.04492}\ } (\bibinfo
  {year} {2018})}\BibitemShut {NoStop}%
\bibitem [{\citenamefont {Quantum}\ \emph {et~al.}(2020)\citenamefont
  {Quantum}, \citenamefont {Collaborators*†}, \citenamefont {Arute},
  \citenamefont {Arya}, \citenamefont {Babbush}, \citenamefont {Bacon},
  \citenamefont {Bardin}, \citenamefont {Barends}, \citenamefont {Boixo},
  \citenamefont {Broughton}, \citenamefont {Buckley} \emph
  {et~al.}}]{google2020}%
  \BibitemOpen
  \bibfield  {author} {\bibinfo {author} {\bibfnamefont {G.~A.}\ \bibnamefont
  {Quantum}}, \bibinfo {author} {\bibnamefont {Collaborators*†}}, \bibinfo
  {author} {\bibfnamefont {F.}~\bibnamefont {Arute}}, \bibinfo {author}
  {\bibfnamefont {K.}~\bibnamefont {Arya}}, \bibinfo {author} {\bibfnamefont
  {R.}~\bibnamefont {Babbush}}, \bibinfo {author} {\bibfnamefont
  {D.}~\bibnamefont {Bacon}}, \bibinfo {author} {\bibfnamefont {J.~C.}\
  \bibnamefont {Bardin}}, \bibinfo {author} {\bibfnamefont {R.}~\bibnamefont
  {Barends}}, \bibinfo {author} {\bibfnamefont {S.}~\bibnamefont {Boixo}},
  \bibinfo {author} {\bibfnamefont {M.}~\bibnamefont {Broughton}}, \bibinfo
  {author} {\bibfnamefont {B.~B.}\ \bibnamefont {Buckley}},  \emph {et~al.},\
  }\href@noop {} {\bibfield  {journal} {\bibinfo  {journal} {Science}\ }\textbf
  {\bibinfo {volume} {369}},\ \bibinfo {pages} {1084} (\bibinfo {year}
  {2020})}\BibitemShut {NoStop}%
\bibitem [{\citenamefont {Heyl}\ \emph {et~al.}(2019)\citenamefont {Heyl},
  \citenamefont {Hauke},\ and\ \citenamefont {Zoller}}]{heyl2019}%
  \BibitemOpen
  \bibfield  {author} {\bibinfo {author} {\bibfnamefont {M.}~\bibnamefont
  {Heyl}}, \bibinfo {author} {\bibfnamefont {P.}~\bibnamefont {Hauke}}, \ and\
  \bibinfo {author} {\bibfnamefont {P.}~\bibnamefont {Zoller}},\ }\href@noop {}
  {\bibfield  {journal} {\bibinfo  {journal} {Science advances}\ }\textbf
  {\bibinfo {volume} {5}},\ \bibinfo {pages} {eaau8342} (\bibinfo {year}
  {2019})}\BibitemShut {NoStop}%
\bibitem [{\citenamefont {Kuper}\ \emph {et~al.}(2022)\citenamefont {Kuper},
  \citenamefont {Pajaud}, \citenamefont {Chinni}, \citenamefont {Poggi},\ and\
  \citenamefont {Jessen}}]{kuper2022}%
  \BibitemOpen
  \bibfield  {author} {\bibinfo {author} {\bibfnamefont {K.~W.}\ \bibnamefont
  {Kuper}}, \bibinfo {author} {\bibfnamefont {J.~P.}\ \bibnamefont {Pajaud}},
  \bibinfo {author} {\bibfnamefont {K.}~\bibnamefont {Chinni}}, \bibinfo
  {author} {\bibfnamefont {P.~M.}\ \bibnamefont {Poggi}}, \ and\ \bibinfo
  {author} {\bibfnamefont {P.~S.}\ \bibnamefont {Jessen}},\ }\href@noop {}
  {\bibfield  {journal} {\bibinfo  {journal} {arXiv:2212.03843}\ } (\bibinfo
  {year} {2022})}\BibitemShut {NoStop}%
\bibitem [{\citenamefont {Gonz\'alez-Garc\'{\i}a}\ \emph
  {et~al.}(2022)\citenamefont {Gonz\'alez-Garc\'{\i}a}, \citenamefont
  {Trivedi},\ and\ \citenamefont {Cirac}}]{GonzalesGarcia2022}%
  \BibitemOpen
  \bibfield  {author} {\bibinfo {author} {\bibfnamefont {G.}~\bibnamefont
  {Gonz\'alez-Garc\'{\i}a}}, \bibinfo {author} {\bibfnamefont {R.}~\bibnamefont
  {Trivedi}}, \ and\ \bibinfo {author} {\bibfnamefont {J.~I.}\ \bibnamefont
  {Cirac}},\ }\href {\doibase 10.1103/PRXQuantum.3.040326} {\bibfield
  {journal} {\bibinfo  {journal} {PRX Quantum}\ }\textbf {\bibinfo {volume}
  {3}},\ \bibinfo {pages} {040326} (\bibinfo {year} {2022})}\BibitemShut
  {NoStop}%
\bibitem [{\citenamefont {Devitt}\ \emph {et~al.}(2013)\citenamefont {Devitt},
  \citenamefont {Munro},\ and\ \citenamefont {Nemoto}}]{devitt2013}%
  \BibitemOpen
  \bibfield  {author} {\bibinfo {author} {\bibfnamefont {S.~J.}\ \bibnamefont
  {Devitt}}, \bibinfo {author} {\bibfnamefont {W.~J.}\ \bibnamefont {Munro}}, \
  and\ \bibinfo {author} {\bibfnamefont {K.}~\bibnamefont {Nemoto}},\
  }\href@noop {} {\bibfield  {journal} {\bibinfo  {journal} {Reports on
  Progress in Physics}\ }\textbf {\bibinfo {volume} {76}},\ \bibinfo {pages}
  {076001} (\bibinfo {year} {2013})}\BibitemShut {NoStop}%
\bibitem [{\citenamefont {Choi}\ \emph {et~al.}(2020)\citenamefont {Choi},
  \citenamefont {Bao}, \citenamefont {Qi},\ and\ \citenamefont
  {Altman}}]{Choi2020}%
  \BibitemOpen
  \bibfield  {author} {\bibinfo {author} {\bibfnamefont {S.}~\bibnamefont
  {Choi}}, \bibinfo {author} {\bibfnamefont {Y.}~\bibnamefont {Bao}}, \bibinfo
  {author} {\bibfnamefont {X.-L.}\ \bibnamefont {Qi}}, \ and\ \bibinfo {author}
  {\bibfnamefont {E.}~\bibnamefont {Altman}},\ }\href {\doibase
  10.1103/PhysRevLett.125.030505} {\bibfield  {journal} {\bibinfo  {journal}
  {Phys. Rev. Lett.}\ }\textbf {\bibinfo {volume} {125}},\ \bibinfo {pages}
  {030505} (\bibinfo {year} {2020})}\BibitemShut {NoStop}%
\bibitem [{\citenamefont {Kueng}\ \emph {et~al.}(2016)\citenamefont {Kueng},
  \citenamefont {Long}, \citenamefont {Doherty},\ and\ \citenamefont
  {Flammia}}]{Kueng2016}%
  \BibitemOpen
  \bibfield  {author} {\bibinfo {author} {\bibfnamefont {R.}~\bibnamefont
  {Kueng}}, \bibinfo {author} {\bibfnamefont {D.~M.}\ \bibnamefont {Long}},
  \bibinfo {author} {\bibfnamefont {A.~C.}\ \bibnamefont {Doherty}}, \ and\
  \bibinfo {author} {\bibfnamefont {S.~T.}\ \bibnamefont {Flammia}},\ }\href
  {\doibase 10.1103/PhysRevLett.117.170502} {\bibfield  {journal} {\bibinfo
  {journal} {Phys. Rev. Lett.}\ }\textbf {\bibinfo {volume} {117}},\ \bibinfo
  {pages} {170502} (\bibinfo {year} {2016})}\BibitemShut {NoStop}%
\bibitem [{\citenamefont {Suzuki}\ \emph {et~al.}(2017)\citenamefont {Suzuki},
  \citenamefont {Fujii},\ and\ \citenamefont {Koashi}}]{Suzuki2017}%
  \BibitemOpen
  \bibfield  {author} {\bibinfo {author} {\bibfnamefont {Y.}~\bibnamefont
  {Suzuki}}, \bibinfo {author} {\bibfnamefont {K.}~\bibnamefont {Fujii}}, \
  and\ \bibinfo {author} {\bibfnamefont {M.}~\bibnamefont {Koashi}},\ }\href
  {\doibase 10.1103/PhysRevLett.119.190503} {\bibfield  {journal} {\bibinfo
  {journal} {Phys. Rev. Lett.}\ }\textbf {\bibinfo {volume} {119}},\ \bibinfo
  {pages} {190503} (\bibinfo {year} {2017})}\BibitemShut {NoStop}%
\bibitem [{\citenamefont {Huang}\ \emph {et~al.}(2019)\citenamefont {Huang},
  \citenamefont {Doherty},\ and\ \citenamefont {Flammia}}]{Huang2019}%
  \BibitemOpen
  \bibfield  {author} {\bibinfo {author} {\bibfnamefont {E.}~\bibnamefont
  {Huang}}, \bibinfo {author} {\bibfnamefont {A.~C.}\ \bibnamefont {Doherty}},
  \ and\ \bibinfo {author} {\bibfnamefont {S.}~\bibnamefont {Flammia}},\ }\href
  {\doibase 10.1103/PhysRevA.99.022313} {\bibfield  {journal} {\bibinfo
  {journal} {Phys. Rev. A}\ }\textbf {\bibinfo {volume} {99}},\ \bibinfo
  {pages} {022313} (\bibinfo {year} {2019})}\BibitemShut {NoStop}%
\bibitem [{\citenamefont {Yan}\ \emph {et~al.}(2020)\citenamefont {Yan},
  \citenamefont {Cincio},\ and\ \citenamefont {Zurek}}]{LS_echo}%
  \BibitemOpen
  \bibfield  {author} {\bibinfo {author} {\bibfnamefont {B.}~\bibnamefont
  {Yan}}, \bibinfo {author} {\bibfnamefont {L.}~\bibnamefont {Cincio}}, \ and\
  \bibinfo {author} {\bibfnamefont {W.~H.}\ \bibnamefont {Zurek}},\ }\href
  {\doibase 10.1103/PhysRevLett.124.160603} {\bibfield  {journal} {\bibinfo
  {journal} {Phys. Rev. Lett.}\ }\textbf {\bibinfo {volume} {124}},\ \bibinfo
  {pages} {160603} (\bibinfo {year} {2020})}\BibitemShut {NoStop}%
\bibitem [{\citenamefont {Khemani}\ \emph {et~al.}(2018)\citenamefont
  {Khemani}, \citenamefont {Vishwanath},\ and\ \citenamefont
  {Huse}}]{operator_spreading}%
  \BibitemOpen
  \bibfield  {author} {\bibinfo {author} {\bibfnamefont {V.}~\bibnamefont
  {Khemani}}, \bibinfo {author} {\bibfnamefont {A.}~\bibnamefont {Vishwanath}},
  \ and\ \bibinfo {author} {\bibfnamefont {D.~A.}\ \bibnamefont {Huse}},\
  }\href {\doibase 10.1103/PhysRevX.8.031057} {\bibfield  {journal} {\bibinfo
  {journal} {Phys. Rev. X}\ }\textbf {\bibinfo {volume} {8}},\ \bibinfo {pages}
  {031057} (\bibinfo {year} {2018})}\BibitemShut {NoStop}%
\bibitem [{\citenamefont {G{\"a}rttner}\ \emph {et~al.}(2017)\citenamefont
  {G{\"a}rttner}, \citenamefont {Bohnet}, \citenamefont {Safavi-Naini},
  \citenamefont {Wall}, \citenamefont {Bollinger},\ and\ \citenamefont
  {Rey}}]{garttner2017}%
  \BibitemOpen
  \bibfield  {author} {\bibinfo {author} {\bibfnamefont {M.}~\bibnamefont
  {G{\"a}rttner}}, \bibinfo {author} {\bibfnamefont {J.~G.}\ \bibnamefont
  {Bohnet}}, \bibinfo {author} {\bibfnamefont {A.}~\bibnamefont
  {Safavi-Naini}}, \bibinfo {author} {\bibfnamefont {M.~L.}\ \bibnamefont
  {Wall}}, \bibinfo {author} {\bibfnamefont {J.~J.}\ \bibnamefont {Bollinger}},
  \ and\ \bibinfo {author} {\bibfnamefont {A.~M.}\ \bibnamefont {Rey}},\
  }\href@noop {} {\bibfield  {journal} {\bibinfo  {journal} {Nature Physics}\
  }\textbf {\bibinfo {volume} {13}},\ \bibinfo {pages} {781} (\bibinfo {year}
  {2017})}\BibitemShut {NoStop}%
\bibitem [{\citenamefont {Garc\'{\i}a-Mata}\ \emph {et~al.}(2018)\citenamefont
  {Garc\'{\i}a-Mata}, \citenamefont {Saraceno}, \citenamefont {Jalabert},
  \citenamefont {Roncaglia},\ and\ \citenamefont {Wisniacki}}]{GarciaMata2018}%
  \BibitemOpen
  \bibfield  {author} {\bibinfo {author} {\bibfnamefont {I.}~\bibnamefont
  {Garc\'{\i}a-Mata}}, \bibinfo {author} {\bibfnamefont {M.}~\bibnamefont
  {Saraceno}}, \bibinfo {author} {\bibfnamefont {R.~A.}\ \bibnamefont
  {Jalabert}}, \bibinfo {author} {\bibfnamefont {A.~J.}\ \bibnamefont
  {Roncaglia}}, \ and\ \bibinfo {author} {\bibfnamefont {D.~A.}\ \bibnamefont
  {Wisniacki}},\ }\href {\doibase 10.1103/PhysRevLett.121.210601} {\bibfield
  {journal} {\bibinfo  {journal} {Phys. Rev. Lett.}\ }\textbf {\bibinfo
  {volume} {121}},\ \bibinfo {pages} {210601} (\bibinfo {year}
  {2018})}\BibitemShut {NoStop}%
\bibitem [{\citenamefont {Shen}\ \emph {et~al.}(2020)\citenamefont {Shen},
  \citenamefont {Zhang}, \citenamefont {You},\ and\ \citenamefont
  {Zhai}}]{Shen2020}%
  \BibitemOpen
  \bibfield  {author} {\bibinfo {author} {\bibfnamefont {H.}~\bibnamefont
  {Shen}}, \bibinfo {author} {\bibfnamefont {P.}~\bibnamefont {Zhang}},
  \bibinfo {author} {\bibfnamefont {Y.-Z.}\ \bibnamefont {You}}, \ and\
  \bibinfo {author} {\bibfnamefont {H.}~\bibnamefont {Zhai}},\ }\href {\doibase
  10.1103/PhysRevLett.124.200504} {\bibfield  {journal} {\bibinfo  {journal}
  {Phys. Rev. Lett.}\ }\textbf {\bibinfo {volume} {124}},\ \bibinfo {pages}
  {200504} (\bibinfo {year} {2020})}\BibitemShut {NoStop}%
\bibitem [{\citenamefont {Jordan}\ and\ \citenamefont
  {Wigner}(1928)}]{jordan1928}%
  \BibitemOpen
  \bibfield  {author} {\bibinfo {author} {\bibfnamefont {P.}~\bibnamefont
  {Jordan}}\ and\ \bibinfo {author} {\bibfnamefont {E.~P.}\ \bibnamefont
  {Wigner}},\ }\href@noop {} {\bibfield  {journal} {\bibinfo  {journal} {Z.
  Phys}\ }\textbf {\bibinfo {volume} {47}},\ \bibinfo {pages} {14} (\bibinfo
  {year} {1928})}\BibitemShut {NoStop}%
\bibitem [{\citenamefont {Tranter}\ \emph {et~al.}(2018)\citenamefont
  {Tranter}, \citenamefont {Love}, \citenamefont {Mintert},\ and\ \citenamefont
  {Coveney}}]{tranter2018}%
  \BibitemOpen
  \bibfield  {author} {\bibinfo {author} {\bibfnamefont {A.}~\bibnamefont
  {Tranter}}, \bibinfo {author} {\bibfnamefont {P.~J.}\ \bibnamefont {Love}},
  \bibinfo {author} {\bibfnamefont {F.}~\bibnamefont {Mintert}}, \ and\
  \bibinfo {author} {\bibfnamefont {P.~V.}\ \bibnamefont {Coveney}},\
  }\href@noop {} {\bibfield  {journal} {\bibinfo  {journal} {J. Chem. Theory
  Comput.}\ }\textbf {\bibinfo {volume} {14}},\ \bibinfo {pages} {5617}
  (\bibinfo {year} {2018})}\BibitemShut {NoStop}%
\bibitem [{\citenamefont {Mi}\ \emph {et~al.}(2021)\citenamefont {Mi},
  \citenamefont {Roushan}, \citenamefont {Quintana}, \citenamefont {Mandra},
  \citenamefont {Marshall}, \citenamefont {Neill}, \citenamefont {Arute},
  \citenamefont {Arya}, \citenamefont {Atalaya}, \citenamefont {Babbush} \emph
  {et~al.}}]{Mi2021}%
  \BibitemOpen
  \bibfield  {author} {\bibinfo {author} {\bibfnamefont {X.}~\bibnamefont
  {Mi}}, \bibinfo {author} {\bibfnamefont {P.}~\bibnamefont {Roushan}},
  \bibinfo {author} {\bibfnamefont {C.}~\bibnamefont {Quintana}}, \bibinfo
  {author} {\bibfnamefont {S.}~\bibnamefont {Mandra}}, \bibinfo {author}
  {\bibfnamefont {J.}~\bibnamefont {Marshall}}, \bibinfo {author}
  {\bibfnamefont {C.}~\bibnamefont {Neill}}, \bibinfo {author} {\bibfnamefont
  {F.}~\bibnamefont {Arute}}, \bibinfo {author} {\bibfnamefont
  {K.}~\bibnamefont {Arya}}, \bibinfo {author} {\bibfnamefont {J.}~\bibnamefont
  {Atalaya}}, \bibinfo {author} {\bibfnamefont {R.}~\bibnamefont {Babbush}},
  \emph {et~al.},\ }\href@noop {} {\bibfield  {journal} {\bibinfo  {journal}
  {Science}\ }\textbf {\bibinfo {volume} {374}},\ \bibinfo {pages} {1479}
  (\bibinfo {year} {2021})}\BibitemShut {NoStop}%
\bibitem [{\citenamefont {Zhang}\ \emph {et~al.}(2024)\citenamefont {Zhang},
  \citenamefont {Kundu}, \citenamefont {Makri}, \citenamefont {Gruebele},\ and\
  \citenamefont {Wolynes}}]{zhang2024}%
  \BibitemOpen
  \bibfield  {author} {\bibinfo {author} {\bibfnamefont {C.}~\bibnamefont
  {Zhang}}, \bibinfo {author} {\bibfnamefont {S.}~\bibnamefont {Kundu}},
  \bibinfo {author} {\bibfnamefont {N.}~\bibnamefont {Makri}}, \bibinfo
  {author} {\bibfnamefont {M.}~\bibnamefont {Gruebele}}, \ and\ \bibinfo
  {author} {\bibfnamefont {P.~G.}\ \bibnamefont {Wolynes}},\ }\href@noop {}
  {\bibfield  {journal} {\bibinfo  {journal} {Proceedings of the National
  Academy of Sciences}\ }\textbf {\bibinfo {volume} {121}},\ \bibinfo {pages}
  {e2321668121} (\bibinfo {year} {2024})}\BibitemShut {NoStop}%
\bibitem [{\citenamefont {Zhang}\ \emph {et~al.}(2022)\citenamefont {Zhang},
  \citenamefont {Wolynes},\ and\ \citenamefont {Gruebele}}]{Zhang2022}%
  \BibitemOpen
  \bibfield  {author} {\bibinfo {author} {\bibfnamefont {C.}~\bibnamefont
  {Zhang}}, \bibinfo {author} {\bibfnamefont {P.~G.}\ \bibnamefont {Wolynes}},
  \ and\ \bibinfo {author} {\bibfnamefont {M.}~\bibnamefont {Gruebele}},\
  }\href {\doibase 10.1103/PhysRevA.105.033322} {\bibfield  {journal} {\bibinfo
   {journal} {Phys. Rev. A}\ }\textbf {\bibinfo {volume} {105}},\ \bibinfo
  {pages} {033322} (\bibinfo {year} {2022})}\BibitemShut {NoStop}%
\bibitem [{\citenamefont {Li}\ \emph {et~al.}(2023)\citenamefont {Li},
  \citenamefont {Halperin}, \citenamefont {Wang},\ and\ \citenamefont
  {Bohn}}]{Li2023}%
  \BibitemOpen
  \bibfield  {author} {\bibinfo {author} {\bibfnamefont {H.}~\bibnamefont
  {Li}}, \bibinfo {author} {\bibfnamefont {E.}~\bibnamefont {Halperin}},
  \bibinfo {author} {\bibfnamefont {R.~R.~W.}\ \bibnamefont {Wang}}, \ and\
  \bibinfo {author} {\bibfnamefont {J.~L.}\ \bibnamefont {Bohn}},\ }\href
  {\doibase 10.1103/PhysRevA.107.032818} {\bibfield  {journal} {\bibinfo
  {journal} {Phys. Rev. A}\ }\textbf {\bibinfo {volume} {107}},\ \bibinfo
  {pages} {032818} (\bibinfo {year} {2023})}\BibitemShut {NoStop}%
\bibitem [{\citenamefont {Sadhasivam}\ \emph {et~al.}(2023)\citenamefont
  {Sadhasivam}, \citenamefont {Meuser}, \citenamefont {Reichman},\ and\
  \citenamefont {Althorpe}}]{Sadhasivam2023}%
  \BibitemOpen
  \bibfield  {author} {\bibinfo {author} {\bibfnamefont {V.~G.}\ \bibnamefont
  {Sadhasivam}}, \bibinfo {author} {\bibfnamefont {L.}~\bibnamefont {Meuser}},
  \bibinfo {author} {\bibfnamefont {D.~R.}\ \bibnamefont {Reichman}}, \ and\
  \bibinfo {author} {\bibfnamefont {S.~C.}\ \bibnamefont {Althorpe}},\
  }\href@noop {} {\bibfield  {journal} {\bibinfo  {journal} {Proceedings of the
  National Academy of Sciences}\ }\textbf {\bibinfo {volume} {120}},\ \bibinfo
  {pages} {e2312378120} (\bibinfo {year} {2023})}\BibitemShut {NoStop}%
\bibitem [{\citenamefont {Miessen}\ \emph {et~al.}(2023)\citenamefont
  {Miessen}, \citenamefont {Ollitrault}, \citenamefont {Tacchino},\ and\
  \citenamefont {Tavernelli}}]{miessen2023}%
  \BibitemOpen
  \bibfield  {author} {\bibinfo {author} {\bibfnamefont {A.}~\bibnamefont
  {Miessen}}, \bibinfo {author} {\bibfnamefont {P.~J.}\ \bibnamefont
  {Ollitrault}}, \bibinfo {author} {\bibfnamefont {F.}~\bibnamefont
  {Tacchino}}, \ and\ \bibinfo {author} {\bibfnamefont {I.}~\bibnamefont
  {Tavernelli}},\ }\href@noop {} {\bibfield  {journal} {\bibinfo  {journal}
  {Nature Computational Science}\ }\textbf {\bibinfo {volume} {3}},\ \bibinfo
  {pages} {25} (\bibinfo {year} {2023})}\BibitemShut {NoStop}%
\bibitem [{\citenamefont {Johnson}(1975)}]{Johnson1975}%
  \BibitemOpen
  \bibfield  {author} {\bibinfo {author} {\bibfnamefont {K.}~\bibnamefont
  {Johnson}},\ }\href@noop {} {\bibfield  {journal} {\bibinfo  {journal}
  {Annual review of physical chemistry}\ }\textbf {\bibinfo {volume} {26}},\
  \bibinfo {pages} {39} (\bibinfo {year} {1975})}\BibitemShut {NoStop}%
\bibitem [{\citenamefont {Shao}\ \emph {et~al.}(2015)\citenamefont {Shao},
  \citenamefont {Gan}, \citenamefont {Epifanovsky}, \citenamefont {Gilbert},
  \citenamefont {Wormit}, \citenamefont {Kussmann}, \citenamefont {Lange},
  \citenamefont {Behn}, \citenamefont {Deng}, \citenamefont {Feng} \emph
  {et~al.}}]{shao2015advances}%
  \BibitemOpen
  \bibfield  {author} {\bibinfo {author} {\bibfnamefont {Y.}~\bibnamefont
  {Shao}}, \bibinfo {author} {\bibfnamefont {Z.}~\bibnamefont {Gan}}, \bibinfo
  {author} {\bibfnamefont {E.}~\bibnamefont {Epifanovsky}}, \bibinfo {author}
  {\bibfnamefont {A.~T.}\ \bibnamefont {Gilbert}}, \bibinfo {author}
  {\bibfnamefont {M.}~\bibnamefont {Wormit}}, \bibinfo {author} {\bibfnamefont
  {J.}~\bibnamefont {Kussmann}}, \bibinfo {author} {\bibfnamefont {A.~W.}\
  \bibnamefont {Lange}}, \bibinfo {author} {\bibfnamefont {A.}~\bibnamefont
  {Behn}}, \bibinfo {author} {\bibfnamefont {J.}~\bibnamefont {Deng}}, \bibinfo
  {author} {\bibfnamefont {X.}~\bibnamefont {Feng}},  \emph {et~al.},\
  }\href@noop {} {\bibfield  {journal} {\bibinfo  {journal} {Molecular
  Physics}\ }\textbf {\bibinfo {volume} {113}},\ \bibinfo {pages} {184}
  (\bibinfo {year} {2015})}\BibitemShut {NoStop}%
\bibitem [{\citenamefont {Lee}\ \emph {et~al.}(2023)\citenamefont {Lee},
  \citenamefont {Lee}, \citenamefont {Zhai}, \citenamefont {Tong},
  \citenamefont {Dalzell}, \citenamefont {Kumar}, \citenamefont {Helms},
  \citenamefont {Gray}, \citenamefont {Cui}, \citenamefont {Liu} \emph
  {et~al.}}]{Lee2023}%
  \BibitemOpen
  \bibfield  {author} {\bibinfo {author} {\bibfnamefont {S.}~\bibnamefont
  {Lee}}, \bibinfo {author} {\bibfnamefont {J.}~\bibnamefont {Lee}}, \bibinfo
  {author} {\bibfnamefont {H.}~\bibnamefont {Zhai}}, \bibinfo {author}
  {\bibfnamefont {Y.}~\bibnamefont {Tong}}, \bibinfo {author} {\bibfnamefont
  {A.~M.}\ \bibnamefont {Dalzell}}, \bibinfo {author} {\bibfnamefont
  {A.}~\bibnamefont {Kumar}}, \bibinfo {author} {\bibfnamefont
  {P.}~\bibnamefont {Helms}}, \bibinfo {author} {\bibfnamefont
  {J.}~\bibnamefont {Gray}}, \bibinfo {author} {\bibfnamefont {Z.-H.}\
  \bibnamefont {Cui}}, \bibinfo {author} {\bibfnamefont {W.}~\bibnamefont
  {Liu}},  \emph {et~al.},\ }\href@noop {} {\bibfield  {journal} {\bibinfo
  {journal} {Nature communications}\ }\textbf {\bibinfo {volume} {14}},\
  \bibinfo {pages} {1952} (\bibinfo {year} {2023})}\BibitemShut {NoStop}%
\bibitem [{\citenamefont {Liu}\ \emph {et~al.}(2023)\citenamefont {Liu},
  \citenamefont {Meitei}, \citenamefont {Chin}, \citenamefont {Dutt},
  \citenamefont {Tao}, \citenamefont {Chuang},\ and\ \citenamefont
  {Van~Voorhis}}]{Liu2023}%
  \BibitemOpen
  \bibfield  {author} {\bibinfo {author} {\bibfnamefont {Y.}~\bibnamefont
  {Liu}}, \bibinfo {author} {\bibfnamefont {O.~R.}\ \bibnamefont {Meitei}},
  \bibinfo {author} {\bibfnamefont {Z.~E.}\ \bibnamefont {Chin}}, \bibinfo
  {author} {\bibfnamefont {A.}~\bibnamefont {Dutt}}, \bibinfo {author}
  {\bibfnamefont {M.}~\bibnamefont {Tao}}, \bibinfo {author} {\bibfnamefont
  {I.~L.}\ \bibnamefont {Chuang}}, \ and\ \bibinfo {author} {\bibfnamefont
  {T.}~\bibnamefont {Van~Voorhis}},\ }\href@noop {} {\bibfield  {journal}
  {\bibinfo  {journal} {J. Chem. Theory Comput.}\ }\textbf {\bibinfo {volume}
  {19}},\ \bibinfo {pages} {2230} (\bibinfo {year} {2023})}\BibitemShut
  {NoStop}%
\bibitem [{\citenamefont {Bergholm}\ \emph {et~al.}(2018)\citenamefont
  {Bergholm}, \citenamefont {Feldt}, \citenamefont {Mckay}, \citenamefont
  {Gidney}, \citenamefont {Woods},\ and\ \citenamefont {Yen}}]{pennylane}%
  \BibitemOpen
  \bibfield  {author} {\bibinfo {author} {\bibfnamefont {V.}~\bibnamefont
  {Bergholm}}, \bibinfo {author} {\bibfnamefont {R.}~\bibnamefont {Feldt}},
  \bibinfo {author} {\bibfnamefont {D.~J.}\ \bibnamefont {Mckay}}, \bibinfo
  {author} {\bibfnamefont {C.}~\bibnamefont {Gidney}}, \bibinfo {author}
  {\bibfnamefont {J.}~\bibnamefont {Woods}}, \ and\ \bibinfo {author}
  {\bibfnamefont {J.}~\bibnamefont {Yen}},\ }\href@noop {} {\bibfield
  {journal} {\bibinfo  {journal} {arXiv:1811.04968}\ } (\bibinfo {year}
  {2018})}\BibitemShut {NoStop}%
\bibitem [{\citenamefont {Tacchino}\ \emph {et~al.}(2020)\citenamefont
  {Tacchino}, \citenamefont {Chiesa}, \citenamefont {Carretta},\ and\
  \citenamefont {Gerace}}]{gate_decomposition}%
  \BibitemOpen
  \bibfield  {author} {\bibinfo {author} {\bibfnamefont {F.}~\bibnamefont
  {Tacchino}}, \bibinfo {author} {\bibfnamefont {A.}~\bibnamefont {Chiesa}},
  \bibinfo {author} {\bibfnamefont {S.}~\bibnamefont {Carretta}}, \ and\
  \bibinfo {author} {\bibfnamefont {D.}~\bibnamefont {Gerace}},\ }\href
  {\doibase 10.1002/qute.201900052} {\bibfield  {journal} {\bibinfo  {journal}
  {Advances in Quantum Technologies}\ }\textbf {\bibinfo {volume} {3}},\
  \bibinfo {pages} {1900052} (\bibinfo {year} {2020})}\BibitemShut {NoStop}%
\bibitem [{\citenamefont {Burrell}\ and\ \citenamefont
  {Osborne}(2007)}]{Burrell2007}%
  \BibitemOpen
  \bibfield  {author} {\bibinfo {author} {\bibfnamefont {C.~K.}\ \bibnamefont
  {Burrell}}\ and\ \bibinfo {author} {\bibfnamefont {T.~J.}\ \bibnamefont
  {Osborne}},\ }\href {\doibase 10.1103/PhysRevLett.99.167201} {\bibfield
  {journal} {\bibinfo  {journal} {Phys. Rev. Lett.}\ }\textbf {\bibinfo
  {volume} {99}},\ \bibinfo {pages} {167201} (\bibinfo {year}
  {2007})}\BibitemShut {NoStop}%
\bibitem [{\citenamefont {Boguslawski}\ \emph {et~al.}(2013)\citenamefont
  {Boguslawski}, \citenamefont {Tecmer}, \citenamefont {Barcza}, \citenamefont
  {Legeza},\ and\ \citenamefont {Reiher}}]{boguslawski2013orbital}%
  \BibitemOpen
  \bibfield  {author} {\bibinfo {author} {\bibfnamefont {K.}~\bibnamefont
  {Boguslawski}}, \bibinfo {author} {\bibfnamefont {P.}~\bibnamefont {Tecmer}},
  \bibinfo {author} {\bibfnamefont {G.}~\bibnamefont {Barcza}}, \bibinfo
  {author} {\bibfnamefont {O.}~\bibnamefont {Legeza}}, \ and\ \bibinfo {author}
  {\bibfnamefont {M.}~\bibnamefont {Reiher}},\ }\href@noop {} {\bibfield
  {journal} {\bibinfo  {journal} {J. Chem. Theory Comput.}\ }\textbf {\bibinfo
  {volume} {9}},\ \bibinfo {pages} {2959} (\bibinfo {year} {2013})}\BibitemShut
  {NoStop}%
\bibitem [{\citenamefont {Hoffmann}(1970)}]{hoffmann1970geometry}%
  \BibitemOpen
  \bibfield  {author} {\bibinfo {author} {\bibfnamefont {R.}~\bibnamefont
  {Hoffmann}},\ }\href@noop {} {\bibfield  {journal} {\bibinfo  {journal} {Pure
  and Applied Chemistry}\ }\textbf {\bibinfo {volume} {24}},\ \bibinfo {pages}
  {567} (\bibinfo {year} {1970})}\BibitemShut {NoStop}%
\bibitem [{\citenamefont {Bauer}\ and\ \citenamefont
  {Nayak}(2013)}]{bauer2013area}%
  \BibitemOpen
  \bibfield  {author} {\bibinfo {author} {\bibfnamefont {B.}~\bibnamefont
  {Bauer}}\ and\ \bibinfo {author} {\bibfnamefont {C.}~\bibnamefont {Nayak}},\
  }\href@noop {} {\bibfield  {journal} {\bibinfo  {journal} {Journal of
  Statistical Mechanics: Theory and Experiment}\ }\textbf {\bibinfo {volume}
  {2013}},\ \bibinfo {pages} {P09005} (\bibinfo {year} {2013})}\BibitemShut
  {NoStop}%
\bibitem [{\citenamefont {Eisert}\ \emph {et~al.}(2010)\citenamefont {Eisert},
  \citenamefont {Cramer},\ and\ \citenamefont {Plenio}}]{area_law}%
  \BibitemOpen
  \bibfield  {author} {\bibinfo {author} {\bibfnamefont {J.}~\bibnamefont
  {Eisert}}, \bibinfo {author} {\bibfnamefont {M.}~\bibnamefont {Cramer}}, \
  and\ \bibinfo {author} {\bibfnamefont {M.~B.}\ \bibnamefont {Plenio}},\
  }\href {\doibase 10.1103/RevModPhys.82.277} {\bibfield  {journal} {\bibinfo
  {journal} {Rev. Mod. Phys.}\ }\textbf {\bibinfo {volume} {82}},\ \bibinfo
  {pages} {277} (\bibinfo {year} {2010})}\BibitemShut {NoStop}%
\bibitem [{\citenamefont {Bertini}\ \emph {et~al.}(2019)\citenamefont
  {Bertini}, \citenamefont {Kos},\ and\ \citenamefont {Prosen}}]{Bertini2019}%
  \BibitemOpen
  \bibfield  {author} {\bibinfo {author} {\bibfnamefont {B.}~\bibnamefont
  {Bertini}}, \bibinfo {author} {\bibfnamefont {P.}~\bibnamefont {Kos}}, \ and\
  \bibinfo {author} {\bibfnamefont {T.~c.~v.}\ \bibnamefont {Prosen}},\ }\href
  {\doibase 10.1103/PhysRevX.9.021033} {\bibfield  {journal} {\bibinfo
  {journal} {Phys. Rev. X}\ }\textbf {\bibinfo {volume} {9}},\ \bibinfo {pages}
  {021033} (\bibinfo {year} {2019})}\BibitemShut {NoStop}%
\bibitem [{\citenamefont {Molina-Esp{\'\i}ritu}\ \emph
  {et~al.}(2015)\citenamefont {Molina-Esp{\'\i}ritu}, \citenamefont {Esquivel},
  \citenamefont {L{\'o}pez-Rosa},\ and\ \citenamefont
  {Dehesa}}]{molina2015quantum}%
  \BibitemOpen
  \bibfield  {author} {\bibinfo {author} {\bibfnamefont {M.}~\bibnamefont
  {Molina-Esp{\'\i}ritu}}, \bibinfo {author} {\bibfnamefont {R.}~\bibnamefont
  {Esquivel}}, \bibinfo {author} {\bibfnamefont {S.}~\bibnamefont
  {L{\'o}pez-Rosa}}, \ and\ \bibinfo {author} {\bibfnamefont {J.}~\bibnamefont
  {Dehesa}},\ }\href@noop {} {\bibfield  {journal} {\bibinfo  {journal} {J.
  Chem. Theory Comput.}\ }\textbf {\bibinfo {volume} {11}},\ \bibinfo {pages}
  {5144} (\bibinfo {year} {2015})}\BibitemShut {NoStop}%
\bibitem [{\citenamefont {Esquivel}\ \emph {et~al.}(2015)\citenamefont
  {Esquivel}, \citenamefont {Molina-Esp{\'\i}ritu}, \citenamefont {Plastino},\
  and\ \citenamefont {Dehesa}}]{esquivel2015quantum}%
  \BibitemOpen
  \bibfield  {author} {\bibinfo {author} {\bibfnamefont {R.~O.}\ \bibnamefont
  {Esquivel}}, \bibinfo {author} {\bibfnamefont {M.}~\bibnamefont
  {Molina-Esp{\'\i}ritu}}, \bibinfo {author} {\bibfnamefont {A.}~\bibnamefont
  {Plastino}}, \ and\ \bibinfo {author} {\bibfnamefont {J.~S.}\ \bibnamefont
  {Dehesa}},\ }\href@noop {} {\bibfield  {journal} {\bibinfo  {journal}
  {International Journal of Quantum Chemistry}\ }\textbf {\bibinfo {volume}
  {115}},\ \bibinfo {pages} {1417} (\bibinfo {year} {2015})}\BibitemShut
  {NoStop}%
\bibitem [{\citenamefont {Ding}\ \emph {et~al.}(2020)\citenamefont {Ding},
  \citenamefont {Mardazad}, \citenamefont {Das}, \citenamefont {Szalay},
  \citenamefont {Schollwöck}, \citenamefont {Zimbor{\'a}s},\ and\
  \citenamefont {Schilling}}]{ding2020concept}%
  \BibitemOpen
  \bibfield  {author} {\bibinfo {author} {\bibfnamefont {L.}~\bibnamefont
  {Ding}}, \bibinfo {author} {\bibfnamefont {S.}~\bibnamefont {Mardazad}},
  \bibinfo {author} {\bibfnamefont {S.}~\bibnamefont {Das}}, \bibinfo {author}
  {\bibfnamefont {S.}~\bibnamefont {Szalay}}, \bibinfo {author} {\bibfnamefont
  {U.}~\bibnamefont {Schollwöck}}, \bibinfo {author} {\bibfnamefont
  {Z.}~\bibnamefont {Zimbor{\'a}s}}, \ and\ \bibinfo {author} {\bibfnamefont
  {C.}~\bibnamefont {Schilling}},\ }\href@noop {} {\bibfield  {journal}
  {\bibinfo  {journal} {J. Chem. Theory Comput.}\ }\textbf {\bibinfo {volume}
  {17}},\ \bibinfo {pages} {79} (\bibinfo {year} {2020})}\BibitemShut {NoStop}%
\end{thebibliography}
\end{document}